      [1999/12/01 v1.4c Il Nuovo Cimento]
\documentclass{cimento}

\usepackage{graphicx}  
\title{Measurement of radiative $\tau$ decays at BaBar}
\author{B.~Oberhof\from{ins:x}\from{ins:y}} 
\instlist{\inst{ins:x} Istituto Nazionale di Fisica Nucleare sezione di Pisa, Pisa, Italy.
\inst{ins:y} Universit\'a di Pisa, Pisa, Italy.}
\PACSes{\PACSit{13.35.Dx} {Decays of Taus} \PACSit{14.60.Fg} {Properties of Taus} \PACSit{01.30.Cc} {Conference Proceedings}}

\begin{document}

\maketitle

\begin{abstract}
We present a study for the measurement of $\tau$ radiative decays, $\tau \rightarrow l \gamma \nu \bar \nu$, 
($ l = e, \mu$) and $\tau \rightarrow \pi \gamma \nu$ using $\sim 500$ fb$^{-1}$ of $e^+$ $e^-$ annihilations recorded 
by the BaBar detector at the PEP-II asymmetric B-factory.
\end{abstract}

\section{Introduction}

Leptonic radiative $\tau$ decays are sensitive to the Lorentz structure of the decay vertex and, in particular, 
to the anomalous magnetic moment of the $\tau$ \cite{ref:sam}; their branching fractions 
as well as the phase space distributions of the decay products have been calculated with high precision in the SM. 
Up to date the only measurement of the branching fraction for $\tau \rightarrow e \gamma \nu \nu$ has been done by the CLEO collaboration \cite{ref:ber}, which quotes the result $(1.75\pm0.06\pm0.17)\times10^{-2}$, 
using 4.68 fb$^{-1}$ of electron positron annihilations. The same experiment 
also quotes the most precise value for the branching fraction for the $\tau \rightarrow \mu \gamma \nu \nu$ 
decay, $(3.61\pm 0.16 \pm 0.35)\times 10^{-3}$. 
Besides the leptonic radiative decays also the hadronic radiative decay $\tau \rightarrow \gamma \pi \nu$ is of 
great interest because its value, which depends on higher order resonance contributions, has been calculated with high precision 
in chiral perturbation theory \cite{ref:zhi} while experimentally it has never been observed. 

The BaBar experiment ran from 2000 to 2008, mainly at the center of mass energy of the $\Upsilon(4S)$, 
with a $\tau^+ \tau^-$ production cross section of $0.919$ nb which, together with the high luminosity, allowed to collect 
roughly 500 million $\tau$-pairs, 2 orders of magnitude more than what done by CLEO. 
The BaBar detector is described in detail elsewhere \cite{ref:bab}.

\section{Analysis Overview}

To search for radiative tau decays we look to events in which the first $\tau$ decays to the selected mode while the 
other one decays to a charged lepton $\tau \rightarrow l \nu \nu$ or to a charged pion and (at most two) optional 
$\pi^0$s, $\tau \rightarrow \pi (n) \pi^0 \nu$, $n=0,1,2$. 
Efficiency calculation and background evaluation have been performed using 
monte carlo events and a complete simulation of the BaBar detector based on GEANT4. 
Generic $\tau$ events are selected mainly by their topology: each event is divided in two hemispheres, 
the signal-side and the tag-side, by the vector which maximizes the sum of the longitudinal 
projections of momenta in the CM frame, i.e. the Thrust. 
Hadronic events tend to have low Thrust values while for QED backgrounds T$\sim$1. 
In the signal side we require exactly one track identified either as an electron, muon or pion and a 
neutral deposit with an energy greater than 50 MeV located between 10 and 50 cm from the track hit on the calorimeter. 
In the tag-side we require either exactly one electron, muon or pion, depending on mode, and a maximum of 5 photon candidates. 
If more than one photon is present in tag side we require a reconstructed $\pi^0$ for every photon pair. 
Since signal photons tend to be collinear to the track we set $\cos \theta_{min}$=0.99 for 
$\tau \rightarrow e \gamma \nu \nu$ and $\cos \theta_{min}$=0.96 for the other two signal modes. 
For $\tau \rightarrow e \gamma \nu \nu$ the selection is complicated by the presence of external bremsstrahlung 
photons which have nearly the same angular distribution as signal photons; for this reason we set an additional 
cut on the minimum invariant mass of the electron-photon system $m_{e \gamma, min}=0.05$ GeV. 
After selection the major background source for $\tau \rightarrow \mu \gamma \nu \nu$ is given by 
non radiative decays associated to a photon from ISR 
while for $\tau \rightarrow e \gamma \nu \nu$ external bremsstrahlung remains the only significant background. 
The situation for the $\tau \rightarrow \pi \gamma \nu$ decay is more complicated because of the lower branching fraction 
and a irreducible background from neutral "split-offs" i.e. clusters in the calorimeter which are caused by 
hadronic interactions between the pion and 
the calorimeter material; an efficient background reduction technique is still under investigation. 
Final efficiencies and purity for the 3 signal modes are reported in table 1. 

\begin{table}[!hb]
  \label{tab:tab}
  \begin{tabular}{rcl}
    \hline
      Decay & $\epsilon$ & S/(S+B)\\
	\hline
      $\tau \rightarrow e \gamma \nu \nu$    & $0.44 \pm 0.01$\% & $0.79 \pm 0.02$\\
	\hline	
      $\tau \rightarrow \mu \gamma \nu \nu$  & $1.72 \pm 0.03$\% & $0.76 \pm 0.03$\\
	\hline
      $\tau \rightarrow \pi \gamma \nu$      & $1.68 \pm 0.05$\% & $0.20 \pm 0.01$\\
    \hline
  \end{tabular}
    \caption{Final selection efficiencies and purity for the three signal modes.}
\end{table}

\section{Conclusions}
This preliminary study shows that the measurement of radiative leptonic decays at BaBar could be done with significant 
improvement with respect to CLEO's measurement 
Further study is necessary to understand if the measurement is suitable to extract useful information on the Lorentz 
structure of the $\tau$ decay vertex and also to understand if and how the systematics for the rare 
decay $\tau \rightarrow \pi \gamma \nu$ can be reduced.

\end{document}